\documentclass[onecolumn,showpacs,aps,prc,nofootinbib,floatfix]{revtex4}

\usepackage{amsmath}
\usepackage{bm}
\usepackage{graphicx}

\usepackage{epsfig}
\newcommand{\beq}{\begin{equation}}
\newcommand{\eeq}{\end{equation}}
\newcommand{\bea}{\vspace{0.25cm}\begin{eqnarray}}
\newcommand{\eea}{\end{eqnarray}}
\newcommand{\ro}{\mbox{{\boldmath
$\rho$}}}

\def\lsim{\mathrel{\rlap{\lower4pt\hbox{\hskip1pt$\sim$}}
    \raise1pt\hbox{$<$}}}         %less than or approx. symbol
\def\gsim{\mathrel{\rlap{\lower4pt\hbox{\hskip1pt$\sim$}}
    \raise1pt\hbox{$>$}}}         %greater than or approx. symbol

\newcommand{\landau}{L.D.~Landau Institute for Theoretical Physics,
        GSP-1, 117940, Kosygina Str. 2, 117334 Moscow, Russia}

\begin{document}

%%%%%%%%%%%%%%%%%%%%%%%%Front Matter%%%%%%%%%%%%%%%%%%%%%%%%%%%%%%%%%%
%%%%%%%%%%%%%%%%%%%%%%%%%%%%%%%%%%%%%%%%%%%%%%%%%%%%%%%%%%%%%%%%%%%%%%
%\renewcommand{\thefootnote}{\fnsymbol{footnote}}

\title{
  Collective nuclear vibrations and initial state shape fluctuations
  in central Pb+Pb collisions: resolving the $v_2$ to $v_3$ puzzle
}
\date{\today}

\author{B.G.~Zakharov}\affiliation{\landau}

\begin{abstract}
We have studied, for the first time,  
the influence of the collective quantum effects in the nuclear wave
functions on the azimuthal anisotropy coefficients $\epsilon_{2,3}$
in the central Pb+Pb collisions at the LHC energies.
With the help of the energy weighted sum rule we demonstrate that the
classical treatment with the Woods-Saxon nuclear density
overestimates the mean square quadrupole moment
of the $^{208}$Pb nucleus by a factor of $\sim 2.2$.
The Monte-Carlo Glauber simulation of the central Pb+Pb collisions
accounting
for the restriction on the quadrupole moment
leads  to
$\epsilon_2/\epsilon_3\approx 0.8$
which allows to resolve the $v_2$-to-$v_3$ puzzle.

\end{abstract}
%
%\pacs{12.38.Mh, 24.85.+p}

\maketitle
%%%%%%%%%%%%%%%%%%%%%%%%%%%%%%%%%%%%%%%%%%%%%%%%%%%%%%%%%%%%%%%%%%%%
\noindent {\bf 1}.
The results of experiments on the heavy ion collisions at RHIC and LHC
give a lot of evidences for formation of the quark-gluon plasma (QGP)
in the initial stage of nuclear collisions.
The hydrodynamic simulations of the hadron production show that
the QGP undergoes early thermalization (at the proper time $\tau_0\sim 0.5-1$
fm)
and flows as an almost ideal fluid
(the ratio of the shear viscosity to the entropy density
is of the order of the theoretical quantum lower limit 
$\eta/s=1/4\pi$) \cite{hydro1,hydro2,hydro3}.
The most effective constraints on the QGP viscosity
come from the hydrodynamic analysis
of the azimuthal dependence of the hadron spectra which is characterized
by the Fourier coefficients $v_n$
\beq
\frac{dN}{d\phi }=\frac{N}{2\pi}
  \left\{1+{\sum_{n=1}^\infty}2v_n\mathrm{cos}\left[n\left(\phi-\Psi_n\right)
  \right]\right\},
\label{eq:10}
\eeq
where $N$ is hadron multiplicity in a certain $p_T$ and rapidity bin,
$\Psi_n$ are the event reaction plane angles.
For smooth initial conditions at midrapidity ($y=0$) in the Fourier
series (\ref{eq:10}) only the terms with $n=2k$ survive.
And the azimuthal anisotropy appears only for noncentral collisions
due to the almond shape of the overlap region of the colliding nuclei
in the transverse plane.
The event plane (for each $n$) in this case coincides with the true
reaction plane
and $\Psi_n=0$.
However, in the presence of fluctuations of the initial QGP entropy
distributions, all the flow coefficients $v_n$
become nonzero and the event plane angles $\Psi_n$ fluctuate around the
true reaction plane. The fluctuations of the initial fireball entropy
distribution is a combined effect of the fluctuations of the nucleon positions
in the colliding nuclei and fluctuations of the entropy production
for a given geometry of the nuclear positions.
The most popular method for evaluation of the initial entropy distribution
for event-by-event simulation of $AA$-collisions is the Monte-Carlo (MC)
wounded nucleon Glauber model \cite{MCGL,GLISS2}.
The even-by-event hydrodynamic
modeling with the MC Glauber (MCG) model initial conditions
has been quite successful in description of
a vast body of experimental data on the flow
coefficients in $AA$-collisions obtained at RHIC and LHC.
However, in the last years it was found that the hydrodynamical
models fail to describe simultaneously $v_2$ and $v_2$ flow coefficients
in the ultra-central ($c\to 0$) Pb+Pb collisions at the LHC energies
\footnote{
Experimentally, the centrality, $c$, of an event
is defined in terms of the charged particle multiplicity.
To a very good accuracy (except for the  most peripheral collisions)
$c$ in terms of the impact parameter $b$ reads
$c=\pi b^2/\sigma_{in}^{AA}$ \cite{centrality}.}.
For central collisions, at $b=0$, the anisotropy of the initial fireball geometry originates completely
from the fluctuations.
The hydrodynamic calculations show \cite{Niemi,vn1} that for small centralities 
in each event the $v_n$ for $n\le 3$ to good accuracy satisfy the
linear response
relation
\beq
v_n\approx k_n\epsilon_n\,,
\label{eq:20}
\eeq
where $\epsilon_n$ are the Fourier coefficients characterizing
the anisotropy of the initial fireball entropy distribution, $\rho_s(\ro)$,
in the transverse plane defined as
\cite{Teaney_en,Ollitraut_en}
\beq
\epsilon_n=\frac{\left|\int d\ro \rho^n e^{in\phi}\rho_s(\ro)\right|}
{\int d\ro \rho^n\rho_s(\ro)}\,.
\label{eq:30}
\eeq
Here it is assumed that the transverse vector $\ro$ is calculated in the 
tranverse c.m. frame, i.e., $\int d\ro \ro\rho_s(\ro)=0$. 
The hydrodynamic calculations give $k_2/k_3>1$, and this ratio
grows with increase of the QGP viscosity. On the other hand, the MCG
calculations  show that  at $b=0$ $\epsilon_2$ and $\epsilon_3$ are 
close to each other (and are $\sim 0.1$ for Pb+Pb collisions).
This leads to prediction that $v_2/v_3>1$. But experimentally
it was observed that  $v_2$ is close to $v_3$
in the ultra-central $2.76$ and $5.02$ TeV Pb+Pb collisions
\cite{CMS_v23,ALICE_v23}.
Since the hydrodynamic prediction for $k_2/k_3$ seems to be very reliable,
this situation looks very puzzling (it is called in the literature
$v_2$-to-$v_2$ puzzle). This leads to a  serious tension for the hydrodynamic
paradigm of heavy ion collisions.

There were several attempts to resolve the $v_2$-to-$v_2$ puzzle
by modifying: the initial conditions \cite{v23_Heinz,v23_Luzum},
the viscosity coefficients \cite{v23_Gale}, and the QGP equation of state of
\cite{v23_EoS}.
However, these attempts have not been successful.
The common feature of all previous analyses devoted to the
$v_2$-to-$v_2$ puzzle is 
the use of the Woods-Saxon (WS) nuclear distribution
for sampling the nucleon positions in the MC simulations
of Pb+Pb collisions.
In fact, this is an universal choice in the physics of
high-energy heavy ion collisions.
However, the MC sampling of nucleon positions with the
WS distribution completely ignores the collective nature
of the long range fluctuations of the nucleon positions.
It is well known that the long range 3D fluctuations of the nuclear
density have a collective nature and are closely related to the
giant nuclear resonances \cite{BM,Greiner}
(for more recent reviews see \cite{Speth,Roca}).
The major vibration mode of the spherical $^{208}$Pb nucleus
corresponds to excitation of the isoscalar giant quadrupole
resonance  \cite{BM,Greiner}. These collective quantum effects are
completely lost if one samples the nuclear configurations
with the WS distribution.
It is clear that an inappropriate description of the 3D long range
fluctuation of the nucleon positions in the colliding nuclei
will translate into incorrect long range fluctuations of the
2D initial fireball entropy density, which are crucial for
$\epsilon_{2,3}$ in the central $AA$-collisions, when
they are driven by fluctuations.

In the present paper we demonstrate that the WS distribution
overestimates considerably the mean square nuclear quadrupole moment
of the $^{208}$Pb nucleus as compared to that obtained
in the quantum treatment of the quadrupole vibrations.
We calculate the azimuthal anisotropy
coefficients $\epsilon_{2,3}$ in Pb+Pb collisions in the MCG
model by sampling the nuclear configurations for ordinary WS distribution
and a modified one 
which reproduces the quantum
mean square nuclear quadrupole moment of the $^{208}$Pb nucleus.
Our results show that for the quantum version the ratio
  $\epsilon_2/\epsilon_3$ becomes substantially smaller than that for
ordinary WS distribution.
The magnitude of the
obtained $\epsilon_2/\epsilon_3$ is small enough to resolve
the $v_2$-to-$v_2$ puzzle.

Note that the ordinary MC simulation is also inadequate
for the isovector dipole mode, which plays an important
role in fluctuations of electromagnetic fields in $AA$-collisions
at the RHIC and LHC energies \cite{Z_e1} (the classical treatment
overestimates the mean square dipole moments for $^{197}$Au
and $^{208}$Pb by a factor of $\sim 5$).
However, the effect
of the isovector dipole vibrations on the entropy fluctuations
and the ration $\epsilon_2/\epsilon_3$
in $AA$-collisions turns out to be very small.
For the first time, the problem with description of the mean square
quadrupole nucleus moments in the ordinary MC simulations with
the WS nuclear distribution and its importance
for the event-by-event analyses of $AA$-collisions was noted in
\cite{Z_e2}.\\

\noindent {\bf 2}.
 We assume that $^{208}$Pb
 nucleus is spherical, and the nuclear density is given by
 the ordinary WS nuclear density
\beq
\rho_{A}(r)=\frac{\rho_0}{1+\exp[(r-R_A)/a]}\,
\label{eq:40}
\eeq
with parameters $R_{A}=(1.12A^{1/3}-0.86/A^{1/3})=6.49$ fm, and
$a=0.54$ fm \cite{GLISS2}. Let us first consider classical calculation of
the nuclear mean square multipole moment. We define
the isoscalar $L$-multipole operator as (see, e.g. \cite{BM,Greiner,Roca})
in terms of the spherical harmonics
\beq
F_L=\sum_{i=1}^A r^L_iY_{Lm}(\hat{\ro}_i)
\label{eq:50}
 \eeq
with $\hat{\ro}_i=\ro/|\ro|$. 
Assuming that the many-body nuclear density factorizes into a product of
the single nucleon WS densities, one can easily obtain (we ignore
a very small effect of the c.m. correlations)
  \beq
  \langle F^{+}_{L}F_L\rangle_{WS}=\frac{A(2L+1)\langle
    r^{2L}\rangle}{4\pi}\,.
  \label{eq:60}
  \eeq
  Of course, this formula becomes invalid in the presences of
  the nucleon correlations.
  Usually, in the MC simulations of $AA$-collisions, the effect of the nuclear
  correlations is included in the approximation of a 
  hard-core repulsion. The short range $NN$-expulsion somewhat suppresses
  the mean square quadrupole moment. But this suppression is not very
  strong.
  More important effect on the multipole moments may come from
  the long range correlations due to quantum collective nuclear excitations.

  The quantum calculation of the mean square quadrupole moment of the $^{208}$Pb
nucleus can be performed with the help of the energy weighted sum
rule (EWSR) (for a review, see \cite{EWSR})
for strength function $S(\omega)$ of the isoscalar quadrupole operator.
For the nuclear ground state the strength function of an operator $F$
reads
\beq
S(\omega)=\sum_n |\langle n|F|0\rangle |\delta(\omega-\omega_n)\,,
\label{eq:70}
  \eeq
  where $\omega_n=E_n-E_0$ and $E_n$ are the energies of the nucleus states.
  The ground state expectation value of the operator $F^{+}F$ can be
  written as
  \beq
  \langle 0 |F^{+}F|0\rangle=m_0\,,
  \label{eq:80}
  \eeq
  where $m_0$ is the zeroth order moment of the strength function.
For an arbitrary $k$ the moment $m_k$ is defined as
  \beq
  m_k= \int_0^{\infty} d\omega \omega^k S(\omega)\,.
  \label{eq:90}
    \eeq
    The ratio $m_1/m_0$ 
    characterizes the typical energy of the states excited by the action
    of the operator $F$ on the ground state, which is usually called the centroid energy $E_c$.
Then, in terms of $E_c$ we can write 
\beq    
\langle 0 |F^{+}F|0\rangle=\frac{m_1}{E_c}\,.
\label{eq:100}
 \eeq
 For the case of interest $F=F_L$ 
 the moment $m_1$ can be evaluated accurately using the EWSR, that 
 gives for $L\ge 2$ \cite{EWSR,Roca}
 \beq
 m_1=\frac{AL(2L+1)^2\langle r^{2L-2}\rangle}{8\pi m_N}\,,
 \label{eq:110}
 \eeq
where $m_N$ is the nucleon mass.
Then from (\ref{eq:60}) and (\ref{eq:100}), we obtain for the ratio of the classical to
the quantum mean square moments
\beq
r=\frac{ \langle 0 |F_L^{+}F_L|0\rangle_{c}}
      {\langle 0 |F_L^{+}F_L|0\rangle_{q}}=
\frac{ 2m_NE_c \langle r^{2L}\rangle}
     {L(2L+1)\langle r^{2L-2}\rangle}\,.
     \label{eq:120}
      \eeq

In the case of the isoscalar
$L=2$ operator the EWSR is exhausted by the isoscalar giant quadrupole
resonance (ISGQR) with $\omega_q\approx 10.89$ MeV and
$\Gamma_q\approx 3$ MeV \cite{IS1}. Calculation
with the Breit-Wigner parametrization of the quadrupole strength
function gives the ISGQR centroid energy
$E_c\approx 11.9$ MeV\footnote{The strength function is proportional
  to the imaginary part of the quadrupole polarisability(susceptibility)
  $\alpha_q$ (see Eq. (20) of \cite{Z_e2}) which satifies the relation
  $\alpha_q(-\omega^*)=\alpha_q^*(\omega)$. For this
  reason there must be used a double Breit-Wigner parametrization
  with the poles at $\pm\omega_q-i\Gamma_q/2$.}.
Using this centroid energy, we obtain from (\ref{eq:120}) for the quadrupole
mode $r\approx 2.2$
\footnote{
  There was an error in the code used in \cite{Z_e2}. 
  This led to underestimating
 the EWSR mean square quadrupole moment by a factor of
  $\sim 3.95$.
For this reason the quadrupole contribution
  in Figs.~2,~3 (dashed lines) of \cite{Z_e2} should be multiplied
  by this factor.}.
This says that the simple probabilistic treatment
of the $^{208}$Pb nucleus with the factorized WS many-body nuclear density
considerably overestimates the 3D-quadrupole fluctuations. One can expect
that this can lead to incorrect predictions
for the 2D-fluctuations of the QGP fireball in the MC simulation
of $AA$-collisions as well.
One of the ways to cure this problem is
to use in the MC sampling of the nucleon positions
the nuclear configurations that have the distribution function
in the square quadrupole moment (we denote it $Q^2$) of the form
\beq
P_{sq}(Q^2)=rP_0(r Q^2)\,,
\label{eq:130}
\eeq
where $P_0$ is the native distribution function of the squared quadrupole
moment for the WS nuclear distribution (i.e. it is calculated without
imposing any filter on the nucleon positions).
The MC sampling of the nucleon positions with the squeezed distribution
$P_{sq}$ automatically guarantees that the colliding nuclei will have
correct mean square quadrupole moments.\\

\noindent {\bf 3}.
We consider the initial condition for the QGP fireball in Pb+Pb collisions
in the
central rapidity region ($y=0$).
For evaluating the distribution of the entropy density in the transverse
plane we use the MCG approach developed in
\cite{Z_gl1,Z_gl2}.
The MCG scheme of \cite{Z_gl1,Z_gl2} allows to perform calculations
describing the nucleon as a one-body state and  
accounting for the meson-baryon component of the physical nucleon.
This model describes very well the data
on the centrality dependence
of the midrapidity charged particle density in $0.2$ TeV Au+Au
collisions at RHIC and $2.76$ TeV Pb+Pb collisions \cite{Z_gl2}.
The theoretical predictions for
$5.02$ TeV Pb+Pb, and 5.44 Xe+Xe collisions \cite{Z_gl3} are also
in very good agreement with the data. 
In the present analysis we perform calculations for the versions
with and without the meson-baryon component.
Both the versions lead to very close predictions for the ratio 
$\epsilon_2\{2\}/\epsilon_3\{2\}$ we are interested in.
Here we briefly sketch the algorithm used in our MCG model for the
version without the meson-baryon component (for this case our model is similar
to the well known MCG generator GLISSANDO \cite{GLISS2}). 
The interested
reader is referred to \cite{Z_gl1,Z_gl2} for the detailed description
of the model and the parameters of the model.

We use two-component scheme \cite{KN} with two kinds of the entropy sources:
corresponding to the wounded nucleons (WN) and to the hard binary collisions
(BC).
The center of each WN source coincides with the position of the WN.
And the center of each BC source is located in the middle between 
the pair of the colliding nucleons.
The suppression of the probability of hard BC
for a given $NN$-interaction
is
controlled by the parameter $\alpha$.
The total event entropy density in the transverse plane
is given by
\beq
\rho_s(\ro)=\sum_{i=1}^{N_{wn}} S_{wn}(\ro-\ro_i)+
\sum_{i=1}^{N_{bc}} S_{bc}(\ro-\ro'_{i})\,,
\label{eq:140}
\eeq
where the $S_{wn}$ terms corresponds to
the sources for wounded constituents and $S_{bc}$ terms
to the binary collisions, $N_{wn}$ and $N_{bc}$ are the number of the WNs
and BCs, respectively.
The entropy distribution for WN and BC sources are written as
\beq
S_{wn}(\ro)=\frac{(1-\alpha)}{2}s(\ro)\,,\,\,\,\,\,
S_{bc}(\ro)=s(\ro)\,,
\label{eq:150}
\eeq
where for
$s(\ro)$ we use a Gaussian distribution
\beq  
s(\ro)=s_0\exp{\left(-\ro^2/\sigma^2\right)}/\pi \sigma^2\,
\label{eq:160}
\eeq
with $s_0$ the total entropy of the source, and $\sigma$ width of the source.  
We perform calculations for
$\sigma= 0.4$ fm. The results
for the anisotropy coefficients 
become sensitive 
to the width of smearing of the sources only for very peripheral 
collisions, and for the central collision they are weakly
sensitive to the values of $\sigma$.

We describe fluctuations of the total  entropy for each source
by the Gamma distribution. The parameters
of the Gamma distribution
have been adjusted to fit the experimental $pp$ data on the mean charged
multiplicity and its variance in the unit pseudorapidity window $|\eta|<0.5$
using the ratio of the entropy to the charged multiplicity
$
dS/dy=C dN_{ch}/d\eta\,,
$
with $C\approx 7.67$ \cite{BM-entropy}.
In the version with the meson-baryon component of our MCG generator
\cite{Z_gl1,Z_gl2} the entropy sources can be produced
in $BB$, $MB$, and $MM$ collisions. Both the versions of the model give
similar predictions for the charged multiplicity. But the optimal
values of the parameter $\alpha$ are somewhat smaller for the
version with the meson-baryon component.
The fit to the data on centrality
dependence of the midrapidity charged particle density 
gives $\alpha\approx 0.09$($0.14$) for the versions with(without) the
meson-baryon component (see \cite{Z_gl2,Z_gl3} for details).

We performed
numerical calculations of the rms
coefficients $\langle \epsilon_n^2\rangle^{1/2}$
(they are usually denoted as $\epsilon_n\{2\}$)
for $n=2$ and $3$ by MC generation of $5\times 10^5$ 
 central ($b=0$) Pb+Pb collisions at $\sqrt{s}=2.76$ and $5.02$ TeV.
 The results for both the versions, with and without  the meson-baryon
 component, are summarized in
table I.
\begin{table}% [!hbt] 
  \begin{tabular}{c|c|c|c|c}
\hline\hline
&
\multicolumn{2}{|c}{\mbox{Pb+Pb}~2.76~TeV} &
\multicolumn{2}{|c}{\mbox{Pb+Pb}~5.02~TeV} \\
\hline
& MC with $P_0(Q^2)$  & MC with $P_{sq}(Q^2)$
& MC with $P_0(Q^2)$  & MC with $P_{sq}(Q^2)$ \\
\hline
$\epsilon_2\{2\}$&0.107(0.112)&0.0946(0.0983)&0.107(0.112)&0.0939(0.0977) \\
$\epsilon_3\{2\}$&0.118(0.121)&0.118(0.121)&0.117(0.12)&0.117(0.12) \\
$\epsilon_2\{2\}/\epsilon_3\{2\}$ &0.907(0.926)&0.802(0.812)&0.915(0.931)&0.802(0.814) \\
\hline
\end{tabular}
  \caption{
    The rms eccentricities $\epsilon_{2,3}\{2\}$
    and the ratio $\epsilon_{2}\{2\}/\epsilon_{3}\{2\}$
    for central $2.76$ and $5.02$ TeV Pb+Pb collisions
    obtained within the MCG model of \cite{Z_gl2} with and without (numbers in brackets)
    the meson-baryon component in the nucleon. For each energy the left
    column shows the results for the sample of nucleon configurations
    without restrictions on the squared quadrupole
    moment $Q^2$ (i.e. for the native distribution $P_0(Q^2)$ for the WS nuclear density),
    and the right one for the sample
    corresponding to the squeezed distribution $P_{sq}(Q^2)$ (see main text
    for details).}
\label{eps1}
\end{table}
From table I one can see that the quantum collective effects for
the quadrupole deformations do not affect $\epsilon_3\{2\}$.
But they give a noticeable reduction of $\epsilon_2\{2\}$.
For the quantum version with the meson-baryon component we obtain
$\epsilon_2\{2\}/\epsilon_3\{2\}\approx0.8$.
For the version without the meson-baryon component we obtain a bit bigger
$\epsilon_{2,3}$. But the change in the ratio $\epsilon_2\{2\}/\epsilon_3\{2\}$
is very small (it is increased by $~\sim 0.01$).

The obtained magnitude of the ratio $\epsilon_2\{2\}/\epsilon_3\{2\}$
allows to resolve the $v_2$-to-$v_3$ puzzle
in the ultra-central Pb+Pb collisions. 
Because the hydrodynamic calculations
give $k_2/k_3\approx 1.2-1.4$
\cite{Olli_k23,v23_Heinz,Olli_Xe,v23_Luzum}
for Pb+Pb collisions at the LHC energies for small centralities ($c\lsim 2$\%).
Then, using our quantum prediction for
$\epsilon_2\{2\}/\epsilon_3\{2\}$
we obtain $v_2/v_3\approx 0.096-1.12$
\footnote{
The numbers in table I were obtained for the factorized WS distribution
without short range $NN$-correlations. We also performed the MCG simulation
with the hard repulsion for the expulsion radius $r_c=0.9$
\cite{nncore} and $0.6$ \cite{HC_HRG} fm. We obtained for
these two versions $\epsilon_2\{2\}/\epsilon_3\{2\}\approx0.845$ and $0.825$,
respectively. These values
also lead to $v_2/v_3$ which is in rather reasonable agreement with the data.
However, one should bear in mind that from the point of view of the
entropy production in $AA$-collisions the real situation
with the contribution to the entropy density
of the short range $NN$-pairs 
may  differ from that in the picture with a big expulsion volume
(as, e.g., in \cite{nncore}).
Say, for a successful dibaryon paradigm of the short range
$NN$-interaction (for reviews, see \cite{dibaryon1,Simonov2})
the expulsion region is not empty, but occupied by a $6q$-cluster.
As in the case of $hD$-scattering \cite{Z_6q}, the $6q$-states can
participate in the color exchanges between the colliding nuclei
and contribute to the entropy generation.
For this reason, in reality the effect of the short range $NN$-configurations
may be of the opposite sign.
}.
This is in reasonable agreement with the ALICE measurements
\cite{ALICE_v23} for 2.76 and 5.02 TeV Pb+Pb collisions that
give in the limit $c\to 0$ $v_2/v_3\approx 1.08\pm 0.05$.

Our calculations have been performed for zero impact parameter $b$.
Due to fluctuations of the multiplicity
(at a given impact parameter), there is some mismatch/smearing between
$b$ and $c$ which experimentally is measured via the multiplicity.
We checked that the effect of this smearing on our
predictions is very small.
Also, to understand the sensitivity of the results to the form
of the squeezed distribution $P_{sq}(Q^2)$ used for filtering
the nucleon configurations in our MCG simulations, we also
performed calculations for sampling
the nuclear configurations with a sharp cutoff in $Q^2$.
The cutoff on the
squared quadrupole moment has been adjusted to
fit the the EWSR mean square quadrupole moment of the $^{208}$Pb nucleus.
This ansatz leads to the value of
$\epsilon_2\{2\}/\epsilon_3\{2\}$ which is in perfect
agreement with that for the ansatz given by (\ref{eq:130}).
This test demonstrates high stability
of our predictions for $\epsilon_2\{2\}/\epsilon_3\{2\}$
against the changes of the $P_{sq}(Q^2)$ distribution.
It means that for $\epsilon_2\{2\}/\epsilon_3\{2\}$
the only crucial quantity is the total mean square quadruple moment
of the colliding nuclei. 

In this preliminary study, we have ignored possible
inadequacy of the MCG simulation with the WS density for the octupole ($L=3$)
vibrations of the $^{208}$Pb nucleus. The mean square octupole moment
can be defined via the EWSR in the same way as for the quadrupole mode
using the ratio of the moments $m_1/m_0$ calculated via the
experimental data on the strength function.
However, for the octupole mode the strength function is not
exhausted by a single resonance, but it
gets contribution from a broad range of $\omega$. It has
peaks at $\omega\sim 2.6$ MeV and $\omega\sim 20$ MeV
(see, e.g. \cite{IS1,Paar,L3}).
From the available experimental data \cite{IS1,Paar,L3} one can conclude
that for $L=3$ the ratio $r$ given by (\ref{eq:120}) is close
to one or a bit smaller. The octupole strength function calculated
in \cite{brown} within the random phase approximation for the Skyrme
interaction also leads to $r\approx 1$. 
However, of course the determination of $r$ from the experimental
data seems to be preferable. But experimental
uncertainties for contribution of the low and high energy
$\omega$-regions
to the EWSR for the octupole mode are rather large. This
renders difficult an accurate calculation of the ratio (\ref{eq:120}).
We checked that the scenario with $r<1$ (for $L=3$) leads
an increase of the
value of $\epsilon_3\{2\}$, and the ratio
$\epsilon_2\{2\}/\epsilon_3\{2\}$ will be somewhat smaller than that obtained
in the present analysis. We leave a detailed MC simulation for this
scenario with accounting for filtering for both the quadrupole and
octupole moments for future work. \\

%%%%%%%%%%%%%%%%%%%%%%%%%%%%%%%%%%%%%%%%%%%%%%%%%%%%%%%%%%%%%%%%%%%%%%%%%

\noindent {\bf 4}.
In summary, we have studied, for the first time,  
the influence of the collective quantum effects in the nuclear wave
functions on the azimuthal anisotropy coefficients $\epsilon_{2,3}$
in the central Pb+Pb collisions at the LHC energies.
We have compared the predictions for the mean square quadrupole moment
of $^{208}$Pb obtained in the
classical probabilistic treatment with the WS 
nuclear distribution with that obtained from the quantum analysis
using the EWSR and the experimental data on the isoscalar giant quadrupole
resonance.
This analysis shows that the classical treatment
overestimates the mean square quadrupole moment
of the $^{208}$Pb nucleus by a factor of $r\approx 2.2$.
In our MCG simulations of Pb+Pb collisions,
we cure this problem 
by sampling the nucleus configurations
for the distribution in the squared quadrupole moment
squeezed by the factor $r$.
This guarantees that the colliding nuclei have the
mean square quadrupole moment predicted by the quantum EWSR.
We have found that the EWSR version of the MCG simulation
leads to a noticeable reduction
of the azimuthal asymmetry $\epsilon_2$, as compared
to the ordinary MC sampling without restrictions
on the quadrupole moments of the colliding nuclei.
The values of $\epsilon_3$ for two versions of the MCG simulations
are practically the same.
For the EWSR version we obtained
$\epsilon_2\{2\}/\epsilon_3\{2\}\approx 0.8$.
This leads to $v_2\{2\}/v_3\{2\}\approx 0.96-1.12$
(if one adopts the hydrodynamic linear response coefficients $k_{2,3}$
from
\cite{Olli_k23,v23_Heinz,Olli_Xe,v23_Luzum}), which is in rather good
agreement with the data from ALICE \cite{ALICE_v23}.

In the present analysis we have addressed only the case of the spherical
$^{208}$Pb nucleus. However, it is clear that for high-energy collisions
of the non-spherical nuclei, like $^{197}$Au+$^{197}$Au
and $^{238}$U+$^{238}$U, the MCG simulations with the
ordinary WS density may be inadequate as well.
This fact may be important for interpretation of the results of
the event shape engineering,
which uses the event multiplicity to select the events with
a certain initial system geometry
(e.g., the tip-tip collisions of the prolate $^{238}$U nuclei
as in the STAR experiment \cite{STAR_UU}).

The quantum collective effects discussed in the present analysis
may be important for analyses of the data on the flow effects in Au+Au
collisions in future experiments at NICA. In the NICA energy region
the critical point effects may influence the medium evolution,
and accurate treatment of the initial state geometry becomes especially
important. \\

\begin{acknowledgments}
I am grateful to S.P.~Kamerdzhiev  for helpful communications on physics
of the giant resonances. I also thank N.N.~Nikolaev for discussing
the results.
This work was partly supported by the RFBR grant 
18-02-40069mega.
\end{acknowledgments}

\end{document}